\documentclass[
12pt,
preprint,preprintnumbers,nofootinbib,
groupedaddress,superscriptaddress,amsmath,amssymb]{revtex4}
\usepackage{graphicx}
\usepackage{dcolumn}
\usepackage{bm}
\usepackage{amssymb}
\usepackage{amsmath}
\usepackage{epsfig}    
\usepackage{color}
\usepackage{slashed}
\usepackage{hhline}
\usepackage{feynmp}
\usepackage{nccmath}

\def\be{\begin{equation}}
\def\ee{\end{equation}}
\newcommand{\bea}{\begin{eqnarray}}
\newcommand{\eea}{\end{eqnarray}}
\newcommand{\nn}{\nonumber}

\numberwithin{equation}{section}
\begin{document}

\title{Neutrinophilic two Higgs doublet model  \\
with $U(1)$ global symmetry }
%

\author{Kingman Cheung}
\email{cheung@phys.nthu.edu.tw}
\affiliation{Physics Division, National Center for Theoretical Sciences, 
Hsinchu, Taiwan 300}
\affiliation{Department of Physics, National Tsing Hua University, 
Hsinchu 300, Taiwan}
\affiliation{Division of Quantum Phases and Devices, School of Physics, 
Konkuk University, Seoul 143-701, Republic of Korea}

\author{Hiroshi Okada}
\email{macokada3hiroshi@cts.nthu.edu.tw}
\affiliation{Physics Division, National Center for Theoretical Sciences, Hsinchu, Taiwan 300}

\author{Yuta Orikasa}
\email{Yuta.Orikasa@utef.cvut.cz}
\affiliation{Institute of Experimental and Applied Physics, Czech Technical University, Prague 12800, Czech Republic}

\date{\today}

\begin{abstract}
We propose a neutrinophilic two-Higgs-doublet model, where the vacuum
expectation value (VEV) of the second Higgs doublet is only induced at
one-loop level via several neutral fermions. Thus, the masses of
active neutrinos arising from the Higgs doublets are naturally small
via such a tiny VEV. We discuss various phenomenology 
of the model, including the neutrino masses and oscillations, 
bounds on non-unitarity, lepton-flavor violations, the 
oblique parameters, the muon anomalous magnetic moment, 
the GeV-scale sterile neutrino candidate arising from the tiny VEV,
and collider signatures.
We finally discuss the possibility of detecting the sterile neutrino 
suggested in the experiment of Future Circular Collider (FCC).
\end{abstract}
\maketitle
\newpage

\section{Introduction}

Two-Higgs-doublet models (THDM) are regarded as the simplest extensions of
the standard model (SM) by adding one more doublet Higgs field to the 
Higgs sector \cite{hunter}. It is the most often studied model
because of its rich phenomenology and accommodation of the Higgs sector of 
supersymmetric models. Nevertheless, THDM's do not have enough 
{matter contents}
to accommodate the small neutrino mass, at least in its simplest versions,
conventionally dubbed as Types I, II, III, and IV.

Here we introduce an additional $U(1)$ global symmetry with a set of 
exotic fermions and a singlet Higgs field, beyond the THDM. 
Among the exotic fermions, there are Dirac and Majorana types.
The first Higgs doublet field $\Phi_1$ is chosen to be 
the SM-like Higgs doublet while the second one $\Phi_2$ to be
inert at tree level. Yet, a tiny vacuum expectation value (VEV) is 
generated at loop level for the second doublet, which is then used to explain
the small neutrino mass. Such a setup can explain the small 
neutrino mass without invoking extremely small Yukawa couplings.

In this work, in addition to showing that the model can explain neutrino mass
and oscillation pattern, and non-unitarity bound, we also show that it 
can be consistent
with existing limits on the lepton-flavor violations and the 
oblique parameters. 
Furthermore, we can have heavier sterile  neutrinos
of mass  ${\cal O}(0.1\sim10)$ GeV, which are induced by the tiny VEV. 
Since the model
also involves some exotic particles at TeV, we briefly describe the
signatures that we can expect at the LHC. 

This paper is organized as follows.
In Sec.~II, we describe the details of the model. 
In Sec.~III, we study the phenomenology
and constraints of the model, in particular, 
the derivations for the formulas of lepton-flavor violations,
muon anomalous magnetic dipole moment ($g-2$), and the oblique parameters.
In Sec.~IV, we present the numerical analysis of the model.
We conclude and discuss in Sec.~V.

\section{ Model setup}

 \begin{widetext}
\begin{center} 
\begin{table}[t]
\begin{tabular}{|c||c|c|c|c|c|c||c|c|c|c|}\hline\hline  
&\multicolumn{6}{c||}{Fermions} & \multicolumn{3}{c|}{Bosons} \\\hline
Fermions& ~$L_L$~ & ~$e_R$~ & ~$L'$ ~ & ~$N_{R_0}$~ & ~$N_1$ ~ 
& ~$N_{R_2}$~ & ~$ \Phi_1$~ & ~$ \Phi_2$~& ~$\varphi$ ~  
\\\hline 
 $SU(2)_L$ & $\bm{2}$  & $\bm{1}$  & $\bm{2}$ & $\bm{1}$ &
 $\bm{1}$  & $\bm{1}$    & $\bm{2}$   & $\bm{2}$  & $\bm{1}$  \\\hline 
$U(1)_Y$ & $-\frac12$ & $-1$  & ${-\frac12}$ & $0$  & $0$ 
 & $0$ & $\frac12$  & $\frac12$ &  $0$   \\\hline
 $U(1)_{}$ & $-1$ & $-1$  & $-\frac15$ & $0$  & $-\frac15$  & $-\frac25$ & $0$ & $\frac35$  & $\frac15$  \\\hline
\end{tabular}
\caption{Field contents of fermions
and their charge assignments under $SU(2)_L\times U(1)_Y\times U(1)_{}$, where each of the flavor index is abbreviated.}
\label{tab:1}
\end{table}
\end{center}
\end{widetext}

In this section, we describe the neutrinophilic model in detail, 
including the bosonic sector, fermion sectors, and the
scalar potential.
First of all, we introduce an additional $U(1)$ global symmetry.
All the fermionic and bosonic contents and their assignments are 
summarized in Table~\ref{tab:1}.
Notice here that the numbers of family for all exotic fermions,
except for $N_{R_0}$ (two families), are three in order to reproduce 
the neutrino oscillation data, and $L'$ and $N_1$ are Dirac-type fermions, 
while $N_{R_0}$ and $N_{R_2}$ are the Majorana fermions. 

For the scalar sector with nonzero VEVs, we introduce two $SU(2)_L$ 
doublet scalars $\Phi_1$ and $\Phi_2$, and an $SU(2)_L$ singlet scalar 
$\varphi$.
Here $\Phi_1$ is supposed to be the SM-like Higgs doublet, while $\Phi_2$  
is supposed to be an inert doublet at tree level.
After spontaneous breaking of $U(1)$ via $\varphi$, the VEV of 
$\Phi_2$ is induced at the one-loop level via exotic fermions.
Thus, a tiny VEV can theoretically be realized, which could be 
natural to generate the tiny neutrino masses.

In the framework of neutrinophilic THDM's, several
scenarios have been considered in literature.  
For example, a tiny VEV is 
induced by bosonic loops at one-loop level with a global $U(1)_{B-L}$
symmetry and thus the active neutrinos are expected to be Dirac fermions
~\cite{Kanemura:2013qva}.
The work in Ref.~\cite{Nomura:2017ezy} had considered a tiny VEV 
generated at bosonic one-loop level with
a $U(1)_R$ gauge symmetry, and all the light SM fermion masses are 
induced via this tiny VEV while the neutrino masses are induced at 
two-loop level as Majorana fermions.
Another work in Ref~\cite{Wang:2016vfj} had considered the scenario in 
neutrinophilic THDM with a $U(1)_L$ global symmetry,
in which neutrino masses are induced at tree-level as Majorana fermions 
and they also discussed the possibility of explaining the 
{anomalous X-ray line.}
\footnote{In the framework of type-II seesaw models, there are
also several models that a small $SU(2)_L$ triplet VEV can be induced at 
loop levels~\cite{Kanemura:2012rj, Okada:2015nca}.}

\subsection{Yukawa interactions and scalar sector}
{\it Yukawa Lagrangian}:
With the current field contents and symmetries, the renormalizable 
Lagrangian in the leptonic sector is given by 
\begin{align}
-{\cal L}_{L}&=(y_\ell)_{ii}\bar L_{L_i}  \Phi_1 e_{R_i} 
+f_{ij}\bar L'_{L_i} \tilde  \Phi_1  N_{R_{1j}} + f'_{ij} \bar N_{L_i} L'_{R_j}  \Phi_1
+g_{ij}\bar L'^C_{R_i}   \Phi_2  N_{R_{2j}} +(y_{L_2})_{i j}\bar L_{L_i} \tilde  \Phi_2 N_{R_{2j}} 
\nn\\&
 + (y_N)_{ij} \bar N_{R_{2i}} N_{L_{1j}} \varphi
 + (y'_N)_{ia} \bar N_{L_{1i}} N_{R_{0 a}} \varphi
 + (y''_N)_{ia} \bar N_{R_{1i}} N^C_{R_{0 a}} \varphi +(M_{D})_{ij} \bar N_{L_{1i}} N_{R_{1j}}\nn\\&
+(M_0)_{aa} \bar N^C_{R_{0a}} N_{R_{0a}}
+(M_{L'})_{ii} \bar L'_{L_i} L'_{R_{i}}
+{\rm c.c.},
\label{eq:lag-lep}
\end{align}
where $(i,j)=1-3$, $a=1,2$, $\tilde \Phi_{1,2} \equiv (i \sigma_2)
\Phi_{1,2}^*$ with $\sigma_2$ being the second Pauli matrix, and the 
mass matrices in the 
last line are diagonal without loss of generality as well as
$y_\ell$.~\footnote{Although $M_D$ is diagonal in general: $9\to3$; we
  select a symmetric $M_D$: $9\to6$; and reduced the parameter $y_{N'}$ by
  three degrees of freedom: $6\to3$. Thus the total degrees of freedom
  is conserved. }

\subsection{Fermion Sector}
First of all, we define the exotic fermion as follows:
\begin{align}
L'_{L(R)}\equiv 
\left[
\begin{array}{c}
N'\\
E'^-
\end{array}\right]_{L(R)},
\end{align}
then the mass eigenvalue of charged fermion is straightforwardly given by $M_{L'}$ in Eq.(\ref{eq:lag-lep}).
The mass matrix for the neutral exotic fermions is a $7\times 7$
block in basis of 
$\Psi\equiv [\nu_L^C, N_{R_0},N_{R_1},N^C_{L_1},N_{R_2},N'_R,N'^C_L]$, and given by
\begin{align}
M_N(\Psi)=
\left[\begin{array}{ccccccc}
{\bf0_{3\times3}} & {\bf0_{3\times2}} & {\bf0_{3\times3}} &{{\bf0_{3\times3}}}  & {m_D} & {\bf0_{3\times3}}  & {\bf0_{3\times3}} \\
{\bf0_{2\times3}} & M_0 & M_{N_3}^T& M_{N_2}^T & {\bf0_{2\times3}} & {\bf0_{2\times3}}  & {\bf0_{2\times3}} \\
{\bf0_{3\times3}} & M_{N_3} & {\bf0_{3\times3}} & M_D^T & {\bf0_{3\times3}} & {\bf0_{3\times3}} & M^T \\
{{\bf0_{3\times3}}} & M_{N_2} & M_D & {\bf0_{3\times3}} & M_{N_1}^T & M' & {\bf0_{3\times3}} \\
{m_D^T} & {\bf0_{3\times2}} & {\bf0_{3\times3}} & M_{N_1} & {\bf0_{3\times3}} & m^T &  {\bf0_{3\times3}} \\
{\bf0_{3\times3}} & {\bf0_{3\times2}} & {\bf0_{3\times3}} & M'^T & m & {\bf0_{3\times3}} &  M_{L'} \\
{\bf0_{3\times3}} & {\bf0_{3\times2}} & M & {\bf0_{3\times3}} & {\bf0_{3\times3}} & M_{L'}^T &  {\bf0_{3\times3}} \\
\end{array}\right],
\label{eq-Nmass}
\end{align}
where we define $m\equiv g_{ij}v_2/\sqrt2$, $m_D\equiv (y_{L_2})_{ij}v_2/\sqrt2$, $M \equiv f_{ij}v_1/\sqrt2$, $M' \equiv f'_{ij}v_1/\sqrt2$, $M_{N_1}\equiv (y_{N})_{ij}v'/\sqrt2$,
$M_{N_2}\equiv (y'_{N})_{ia}v'/\sqrt2$, $M_{N_3}\equiv (y''_{N})_{ia}v'/\sqrt2$.
Then this matrix is diagonalized by a $20 \times 20$ unitary matrix 
$V_N$ as $M_{\psi_\alpha}\equiv (V_N M_N V_N^T)_\alpha$ $(\alpha=1\sim20)$, 
where $M_{\psi_\alpha}$ consists of the mass eigenvalues.

\subsection{ Scalar potential}
\label{subsec-sp}
In our model, the scalar potential is given by
\begin{align}
V = &
\mu_{\varphi}^2 |\varphi|^2 + \lambda_{\varphi} |\varphi|^4 + \sum_{i =1,2} \lambda_{\varphi \Phi_i} |\varphi|^2 |\Phi_i|^2  \nn \\
& + \mu_{11}^2 |\Phi_1|^2 + \mu_{22}^2 |\Phi_2|^2  + \frac{\lambda_1}{2} |\Phi_1|^4 + \frac{\lambda_2}{2} |\Phi_2|^4 + \lambda_3 |\Phi_1|^2 |\Phi_2|^2 
+ \lambda_4 |\Phi_1^\dagger \Phi_2|^2
\label{eq:lag-pot-2},
\end{align}
where we have chosen some parameters in the potential such that $\langle
\Phi_2 \rangle \equiv v_2/\sqrt{2}=0$ at tree level, while
    {$\varphi=\frac{1}{\sqrt2}(v' + \varphi_{R} + i G)$} with $
    v'\neq0$ and $\langle \Phi_1 \rangle \equiv
    v_1/\sqrt{2}\neq0$. Here $\varphi_R$ is assumed to be the mass
    eigenstate that suggests the mass of $\varphi_R$ is larger than
    the other mass eigenvalues.  $G$ is the physical Goldstone boson
    (GB) that does not mix with other particles.
    To achieve the inert $\Phi_2$,
    we impose the inert conditions as follows:
 \[0< \lambda_2,\quad
  0\le 2\mu_{22}^2+(\lambda_3+\lambda_4)v_1^2+\lambda_{\varphi\Phi_2} v'^2. 
\]
\begin{figure}[t]
\begin{center}
\includegraphics[width=5cm]{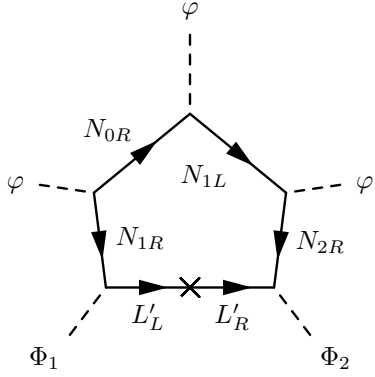} 
\caption{1 loop induced 5 dimensional operator}
\label{fig:1loop}
\end{center}
\end{figure}
A five-demensional operator $\lambda_{5}\varphi^3 \Phi_2^\dagger \Phi_1$ 
can be generated at one-loop level as shown in Fig.\ref{fig:1loop}. 
The formula is given by 
\begin{align}
\lambda_5&=
\sum_{a=1}^{20}
\frac{4 M_{L'_j}}{\left(4 \pi\right)^2}
\left[f_{i, m} g_{i, j} \left(y_N\right)_{j, k} \left(y'_N\right)_{k, l} \left(y''_N\right)_{m, l} \right]
F\left(4, 1, \{ M_{0_i}^2, M_{L'_j}^2, M_{D_k}^2, M_{D_l}^2\}\right), 
\end{align}
where the explicit expression for 
$F\left(4, 1, \{ M_{0_i}, M_{L'_j}, M_{D_k}, M_{D_l}\}\right)$ is given 
in Appendix A.
After the spontaneous symmetry breaking, an effective mass term $\mu_{12}^2\Phi^\dag_2\Phi_1=\frac{\lambda_{5} v'^{3}}{2 \sqrt{2}} \Phi_2^\dagger \Phi_1$ is 
obtained.
The resultant scalar potential in the THDM Higgs sector is given by {
\begin{align}
V_{THDM}=& \mu_{12}^2(\Phi_1^\dag\Phi_2+{\rm c.c.}) + \mu_{11}'^2 |\Phi_1|^2 + \mu_{22}'^2 |\Phi_2|^2  \nn \\
&+ \frac{\lambda_1}{2} |\Phi_1|^4 + \frac{\lambda_2}{2} |\Phi_2|^4 + \lambda_3 |\Phi_1|^2 |\Phi_2|^2 
+ \lambda_4 |\Phi_1^\dagger \Phi_2|^2
 \label{eq:lag-effpot},
\end{align} }
where $\mu_{11(22)}'^2\equiv \mu_{11(22)}^2+\lambda_{\varphi\Phi_{1(2)}}v'^2/2$,
$\langle\Phi_i\rangle\equiv v_i/\sqrt2$ ($i=1-2$) and we choose 
$\mu_{12}^2$ to be negative, while $\mu_{11}^2$ to be positive, and assume
that $v_2/v' \ll 1$.
Taking $v_2/v_1 \ll 1$, 
we finally obtain the formula for the VEV of $\Phi_2$ as
\begin{align}
v_2\approx\frac{2 v_1 \mu_{12}^2}{2 \mu_{22}'^2+v_1^2(\lambda_{3}+\lambda_{4}) }.
\end{align}

Including their VEVs, the scalar fields are parameterized as 
\begin{align}
&\Phi_1 =\left[
\begin{array}{c}
h_1^+\\
\frac{v_1+h_1 + i a_1}{\sqrt2}   
\end{array}\right],\quad 
\Phi_2 =\left[
\begin{array}{c}
h_2^+\\
\frac{v_2+h_2 + i a_2}{\sqrt2}   
\end{array}\right]. 
\label{component}
\end{align}
After the spontaneous symmetry breaking, the neutral scalar bosons 
{$\varphi_R$ and $h_1$ mix each other to form mass eigenstates. Note 
that the VEV of the second Higgs doublet is too small for a sizeable
mixing with $h_2$. The pseudoscalar components and the charged components
are rotated to give the zero-mass Goldstone bosons and the physical
pseudoscalar Higgs boson and charged Higgs boson, respectively. 
They are given in the following expressions:}
\begin{align} 
\label{eq:diagonalize}
& \text{Diag.}(m_{H_1^0}^2, m_{H_2^0}^2)= O_ H m^2( \varphi_R, h_1) O_H^T, \nonumber \\
& \text{Diag.}(m_{Z^0}^2,m_{A^0}^2)= O_ I m^2 (a_1, a_2) O_I^T, \nonumber \\
& \text{Diag.}(m_{\omega^\pm}^2, m_{H^\pm}^2)= O_C m^2(h_1^\pm,h_2^\pm) O_C^T,
& \end{align}
where $O_{H, C, I}$ denotes the mixing matrices which diagonalize the mass matrices accordingly.
{Here $Z^0$ and $\omega^\pm$ are zero-mass Goldstone bosons 
to be absorbed as the longitudinal component of 
the neutral SM gauge boson $Z$ and charged gauge boson $W^\pm$ respectively.}
The mass matrices in the right-hand side
of Eq.~(\ref{eq:diagonalize}) are given by the parameters in the 
scalar potential.
For neutral CP-even components we obtain
\begin{align}
&m^2(\varphi_R, h_1)\sim
\left[
\begin{array}{cc}
2 \lambda_\varphi v'^2 & v_1 v' \lambda_{\varphi \Phi_1}  \\ 
v_1 v' \lambda_{\varphi \Phi_1}  & 2\lambda_1 v_1^2 - \frac{v_2 \mu^2_{12}}{v_1}  \\
\end{array}\right], 
\end{align}
where $H_{1}^0(\equiv h_{SM})$ is the SM-like Higgs in our notation, and $h_2$ does not mix in the limit of $\mu_{12}, v_2 \ll v_1,v_\varphi$;
$m_{h_2}^2 \sim2\lambda_2 v_2^2 - \frac{v_2 \mu^2_{12}}{v_2}$.
We also obtain the mass matrices for CP-odd and charged components as 
\begin{align}
&m^2(a_1,a_2)=
\left[
\begin{array}{cc}
-\frac{v_2 \mu_{12}^2}{v_1} & \mu_{12}^2\\ 
 \mu_{12}^2 & -\frac{v_1 \mu_{12}^2}{v_2} \\
\end{array}\right],\ 
 m_{A^0}^2 = -\frac{(v_1^2+v_2^2) \mu_{12}^2}{v_1 v_2},\\
&m^2(h_1^\pm,h_2^\pm)=
\left[
\begin{array}{cc}
-\frac{v_2(\lambda_{4} v_1 v_2 +2\mu_{12}^2)}{2v_1} &
\frac{ \lambda_{4} v_1 v_2}{2} + \mu_{12}^2\\ 
\frac{ \lambda_{4} v_1 v_2}{2} + \mu_{12}^2 & 
-\frac{v_1( \lambda_{4} v_1 v_2  +2\mu_{12}^2)}{2v_2} \\
\end{array}\right],\ 
 m_{H^\pm}^2 = -\frac{(v_1^2+v_2^2)( \lambda_{4}v_1 v_2 + 2\mu_{12}^2)}{2v_1 v_2}.
 \end{align}
Here we explicitly show the $2\times 2$ matrices; $O_R, O_C,O_I$, as 
\begin{align}
 O_ R
&\equiv
\left[
\begin{array}{cc}
c_\alpha & s_\alpha \\ 
-s_\alpha & c_\alpha \\
\end{array}\right],\quad
\quad
O_ I
\equiv 
O_C=
\left[
\begin{array}{cc}
c_\beta & s_\beta \\ 
-s_\beta & c_\beta \\
\end{array}\right],\quad
s_\beta=\frac{v_2}{\sqrt{v_1^2+v_2^2}},
\end{align}
where $c_{\alpha(\beta)}\equiv \cos \alpha(\beta)$ and
$s_\alpha(\beta)\equiv \sin \alpha(\beta)$, and we define $v\equiv
\sqrt{v_1^2 + v_2^2}$, $\tan\beta\equiv \frac{v_2}{v_1}$ which lead
$v_1=v\cos\beta$ and $v_2=v\sin\beta$ as in the other THDMs.  {\it
  Since $v_2 \ll v_1$ is achieved theoretically, $s_\beta \ll 1$ is
  realized.}  Also $s_\alpha$ is written in terms of the elements of
{
$m^2(\varphi_R ,h_1)$,} which is restricted by the current 
experimental data at LHC $s_\alpha\lesssim 0.3$.
Note that there is an advantage of introducing fermions 
inside the loop instead of bosons~\cite{Kanemura:2013qva, Nomura:2017ezy},
because of the positivity of the fermion-loop contributions to
the pure quartic couplings.
Hence the vacuum stability can easily be realized~\cite{Cheung:2016ypw}.

\section{Phenomenology and Constraints}
 
\subsection{Neutrino masses and Oscillations}

The charged-lepton mass is given by $m_\ell =y_\ell v/\sqrt2$
after the electroweak symmetry breaking, where $m_\ell$ is assumed to
be the mass eigenstate.
Let us redefine the neutral mass matrix $M_N$, its mixing matrix 
$V_N$ and mass eigenvalues $M_{\psi}$ as two by two block-mass matrices 
for the convenience of discussing the non-unitarity of leptonic
mixing matrix~\cite{Agostinho:2017wfs}:
  \begin{align}
M_N & =
\left[\begin{array}{cc} 
0_{3\times 3} & m_{3\times 17}  \\
m^T_{17\times 3} &M_{17\times 17} \\
  \end{array}\right],\quad 
 M_\psi  =
\left[\begin{array}{cc} 
d_{3\times 3} & 0_{3\times 17}  \\
0^T_{17\times 3} &D_{17\times 17} \\
  \end{array}\right],\\
  V_N&=
\left[\begin{array}{cc} 
(V_N)_{3\times 3} & 0_{3\times 17}  \\
0^T_{17\times 3} &(V_N)_{17\times 17} \\
  \end{array}\right]
\left[\begin{array}{cc} 
1_{3\times 3} & X^\dag_{3\times 17}  \\
X_{17\times 3} &1_{17\times 17} \\
  \end{array}\right]  
  ,\quad 
  \end{align}
where $(V_N)_{3\times 3}$ and $d_{3\times3}$ correspond, respectively,
to the lepton-mixing matrix with non-unitarity,
 and mass eigenvalues of active neutrinos.
 With several steps, $X$ can be parametrized by
 \begin{equation}
 X=\pm i\sqrt{D^{-1}} {\cal O} \sqrt{d},
 \end{equation}
where ${\cal O}$ is an arbitrary $17\times 3$ matrix with 45
degrees of freedom, satisfying ${\cal O}^T{\cal O}=1_{3\times3}$ but 
${\cal O}{\cal O}^T\neq 1_{17\times17}$. 
Next, consider the Hermitian matrix $X^\dag X$ being diagonalized by a 
unitary $3\times 3$ mixing matrix $U$, i.e., $d_X^2 \equiv U^\dag X^\dag X U$. 
Then the non-unitarity parameter $\eta$, which is defined by 
$(V_N)_{3\times 3}\equiv (1-\eta) V_{MNS}$, should be smaller than the following
bounds that arise from global constraints in 
Ref.~\cite{Fernandez-Martinez:2016lgt}
 \begin{align}
 &|2\eta| \simeq |V_k d_X^2 V^\dag_k|\lesssim
 \left[\begin{array}{ccc} 
2.5\times 10^{-3} & 2.4\times 10^{-5} & 2.7\times 10^{-3} \\
2.4\times 10^{-5} & 4.0\times 10^{-4} &  1.2\times 10^{-3} \\
2.7\times 10^{-3} & 1.2\times 10^{-3} & 5.6\times 10^{-3} \\
  \end{array}\right],\\
&V_{MNS}=
\left[\begin{array}{ccc} {c_{13}}c_{12} &c_{13}s_{12} & s_{13} e^{-i\delta}\\
 -c_{23}s_{12}-s_{23}s_{13}c_{12}e^{i\delta} & c_{23}c_{12}-s_{23}s_{13}s_{12}e^{i\delta} & s_{23}c_{13}\\
  s_{23}s_{12}-c_{23}s_{13}c_{12}e^{i\delta} & -s_{23}c_{12}-c_{23}s_{13}s_{12}e^{i\delta} & c_{23}c_{13}\\
  \end{array}\right]
\left[\begin{array}{ccc} e^{i\alpha_1/2} & 0 &0 \\
0 & e^{i\alpha_2/2} & 0 \\
0 & 0 &  1\\
  \end{array}\right],
 \end{align}
where $V_{MNS}$ is the unitary $3\times 3$ lepton-mixing matrix that 
is observed, and $V_k\equiv (V_N)_{3\times 3} U_{3\times 3} 
(\sqrt{1+d_X^2})_{3\times 3}$. 
In our numerical analysis, we implicitly satisfy this condition.
~\footnote{
This can easily be satisfied by controlling 45 free parameters of ${\cal O}$.}

In addition to the bounds on non-unitarity, we further impose 
 the following ranges on
\begin{align}
 V^\dag_{MNS} d V_{MNS}^*
= \left[\begin{array}{ccc} 
0.0845-0.475 & 0.0629-0.971 &0.0411-0.964 \\* & 1.44-3.49 &  1.94-2.85 \\* & * &   1.22-3.33\\
  \end{array}\right]\times 10^{-11}\ {\rm GeV},\label{eq:exp-Neutmass-NH}
  \end{align}
where we have used the following neutrino oscillation data at 
$3\sigma$~\cite{Forero:2014bxa} in case of normal hierarchy (NH) 
given by~\footnote{
Recently $\delta=-\pi/2$ is experimentally favored. But our result 
does not change significantly, even if we fix $\delta=-\pi/2$.}
\begin{align}
& 0.278 \leq s_{12}^2 \leq 0.375, \
 0.392 \leq s_{23}^2 \leq 0.643, \
 0.0177 \leq s_{13}^2 \leq 0.0294,  \  \delta\in [-\pi,\pi],
\nn \\
& 
  \sqrt{m_{\nu_3}^2 -\frac{m_{\nu_1}^2+m_{\nu_2}^2}2} =(\sqrt{23.0}-\sqrt{26.5}) \times10^{-11} \ {\rm GeV},  \\
  & \sqrt{m_{\nu_2}^2 -  m_{\nu_1}^2} =(\sqrt{0.711}-\sqrt{0.818}) \times10^{-11} \ {\rm GeV}, 
   \label{eq:neut-exp}
  \end{align}
and the Majorana phases $\alpha_{1,2}$ taken to be 
$\alpha_{1,2}\in[-\pi,\pi]$.\\

In case of inverted hierarchy (IH) we also impose  the following ranges 
at 3$\sigma$ confidential level~\cite{Forero:2014bxa}:
\begin{align}
 V^\dag_{MNS} d V_{MNS}^*
 &=
 \left[\begin{array}{ccc} 
1.00-5.00 & 0.00237-3.83 &0.00256-3.94 \\
* & 0.00279-3.08 &  0.365-2.60 \\
* & * &   0.00500-3.30\\
  \end{array}\right]\times 10^{-11}\ {\rm GeV},\label{eq:exp-Neutmass-IH}\\
&
 0.403 \leq s_{23}^2 \leq 0.640, \
 0.0183 \leq s_{13}^2 \leq 0.0297, 
\nn \\
&   \sqrt{\frac{m_{\nu_1}^2+m_{\nu_2}^2}2-m_{\nu_3}^2} =(\sqrt{22.0}-\sqrt{25.4}) \times10^{-11} \ {\rm GeV}, 
  \end{align}
 where the other values are same as the case of NH.
  

\begin{table}[t]
\begin{tabular}{c|c|c|c} \hline
Process & $(i,j)$ & Experimental bounds ($90\%$ CL) & References \\ \hline
$\mu^{-} \to e^{-} \gamma$ & $(2,1)$ &
	${BR}(\mu \to e\gamma) < 4.2 \times 10^{-13}$ & \cite{TheMEG:2016wtm} \\
$\tau^{-} \to e^{-} \gamma$ & $(3,1)$ &
	${Br}(\tau \to e\gamma) < 3.3 \times 10^{-8}$ & \cite{Adam:2013mnn} \\
$\tau^{-} \to \mu^{-} \gamma$ & $(3,2)$ &
	${BR}(\tau \to \mu\gamma) < 4.4 \times 10^{-8}$ & \cite{Adam:2013mnn}   \\ \hline
\end{tabular}
\caption{Summary of $\ell_i \to \ell_j \gamma$ process and the lower bound of experimental data.}
\label{tab:Cif}
\end{table}

\subsection{Sterile neutrino}
Due to two of the blocks in the mass matrix for neutral fermions,
$m$ and $m_D$,  in Eq.~(\ref{eq-Nmass}),
we can have another three lighter {neutral} fermions in addition to 
the {three} active neutrinos.
Hence the model can provide 
GeV-scale sterile neutrino(s) that have been studied in the 
Future Circular Collider (FCC) 
proposal~\cite{Blondel:2014bra, Alekhin:2015byh}.
Here let us focus on the lightest sterile fermion $\psi_4\equiv \nu_s$, 
and its mass is defined by $m_{\nu_s}$. 
Since the testability of FCC is provided in terms of $m_{\nu_s}$ and its mixing between $\nu_s$ and three active neutrinos~\cite{Rasmussen:2016njh},
we define its mixing as follows:
\begin{align} 
\theta_s\equiv \sqrt{|V_{4,1}|^2+|V_{4,2}|^2+|V_{4,3}|^2},
\end{align}
where $\theta_s$ depends on each of mass values in Eq.~(\ref{eq-Nmass}).
\footnote{One may consider the possibility of a (decaying) dark matter
  candidate with a lighter mass scale of keV or MeV, since single
  photon emission can be possible due to the mixing whose form is the
  same as the sterile one. However, since the typical mixing of our
  model at this mass scale is 0.01$\sim$0.0001, which is too large to
  explain, {\it e.g.}, x-ray line at 3.55 keV or 511 keV line, which
  requires a typical mixing $10^{-5}\sim 10^{-6}$. Thus, the only
  possibility to detect in experiments could be sterile neutrinos.}  The
concrete analysis will be give in the next section.

\subsection{Lepton Flavor Violations (LFVs)}
First of all, we rewrite the leptonic interacting Lagrangian in terms of 
the mass eigenstates as follows:
\begin{align}
-{\cal L}_{\text int}^L&=
-c_\beta  \sum_{i, j=1}^{3}\sum_{a=1}^{20} (y_{L_2})_{i,j} V^T_{N_{j+11,a}} \bar \ell_{L_i} \psi_{R_a} H^-+{\text h.c.}.
\end{align}
Then lepton-flavor violating processes  {\it $\ell_i\to\ell_j\gamma$} 
will give constraints on our parameters, where the experimental bounds 
are listed in Table.~\ref{tab:Cif}. The branching ratio for 
$\ell_i\to\ell_j\gamma$ is given by
 \begin{align}
& BR(\ell_i\to\ell_j\gamma)
 \approx \frac{48\pi^3\alpha_{em} C_{ij} c_\beta^2 }{G_F^2}
\left|\sum_{k,k'=1}^{3} \sum_{\alpha=1}^{20}
\frac{(y_{L_2})_{j,k} V^T_{N_{k+11,\alpha}} (y_{L_2}^\dag)_{k', i} 
V^*_{N_{\alpha,k'+11}}
}{(4\pi)^2}  F_{lfv}(M_{\psi_\alpha},m_{H^\pm})\right|^2\nn\\
& \approx \frac{192\pi^3\alpha_{em} C_{ij} }{(4\pi)^2 v_2^4 G_F^2}
\left|\sum_{k,k'=1}^{3} \sum_{\alpha=1}^{20}
{m_{D_{j,k}} V^T_{N_{k+11,\alpha}} m_{D_{k', i}}^\dag 
V^*_{N_{\alpha,k'+11}} 
} 
 F_{lfv}(M_{\psi_\alpha},m_{H^\pm})\right|^2,\\
& F_{lfv}(m_a,m_b)=\frac{2 m_a^6 +3 m_a^4 m_b^2 -6 m_a^2 m_b^4 +m_b^6+12 m_a^4 m_b^2 \ln(m_b/m_a)}{12(m_a^2-m_b^2)^4},
 \end{align}
 where $\alpha_{em}\approx1/137$ is the fine-structure constant,
$C_{ij}=(1,0.178,0.174)$ for ($(i,j)=((2,1),(3,2),(3,1)$), ${G_F}\approx1.17\times 10^{-5}$ GeV$^{-2}$ is the Fermi constant.

{\it Muon anomalous magnetic dipole moment $(g-2)_{\mu}$}:
Through the same process as the above LFVs, there exists the 
contribution to $(g-2)_{\mu}$, and 
its form $\Delta a_\mu$ is simply given by
\begin{align}
\Delta a_\mu \approx -\frac{m_\mu^2 c_\beta^2} {(4\pi)^2}
\left[\sum_{k,k'=1}^{3} \sum_{\alpha=1}^{20}
 {(y_{L_2})_{2,k} V^T_{N_{k+11,\alpha}} (y_{L_2}^\dag)_{k', 2} V^*_{N_{\alpha,k'+11}} 
 }\right]  F_{lfv}(M_{\psi_\alpha},m_{H^\pm})\nn\\
 \approx -\frac{2 m_\mu^2} {(4\pi)^2 v_2^2}
\left[ \sum_{k,k'=1}^{3} \sum_{\alpha=1}^{20}
 {m_{D_{2,k}} V^T_{N_{k+11,\alpha}} m_{D_{k', 2}}^\dag V^*_{N_{\alpha,k'+11} }
 }\right]  F_{lfv}(M_{\psi_\alpha},m_{H^\pm}).
\end{align}
Although this value can be tested by current experiments $\Delta
a_\mu=(28.8\pm8.0)\times10^{-10}$~\cite{Agashe:2014kda}, one cannot
obtain a positive muon $g-2$ {in the current model.}

\subsection{Oblique parameters} 
\label{subseq.st}
Since we have exotic fermions $L'$ with $SU(2)_L$ doublet,
we have to consider the oblique parameters that restrict the 
mass hierarchy between each of the components of multiple fermions.
In our case, the masses between $E'$ and $\psi_a$ are restricted.
The first task is to write down their kinetically interacting Lagrangians in terms of  mass eigenstate,
and they are give by
\begin{align}
{\cal L}&\sim\frac{g_2}{\sqrt2}\sum_{a=1}^{20}\left(V_{N_{a,\alpha+17}} \bar\psi_a\gamma^\mu P_L E'_\alpha W^+_\mu+V^*_{N_{a,\alpha+14}} \bar\psi_a\gamma^\mu P_R E'_\alpha W^+_\mu+{\rm h.c.}\right)\\
&+\frac{g_2}{2c_w}\sum_{a,b=1}^{20}
\left[ V_{N_{a,\alpha+17}} V^\dag_{N_{\alpha+17,b}} \bar\psi_a\gamma^\mu P_L\psi_b
+ V^*_{N_{a,\alpha+14}} V^T_{N_{\alpha+14,b}} \bar\psi_a\gamma^\mu P_R\psi_b
+\left(-1+2s_w^2\right)\bar E'_\alpha\gamma^\mu  E'_\alpha \right] Z_\mu,
\end{align}
Here we focus on the new physics contributions to $\Delta S$ and $\Delta T$ 
parameters in the case $\Delta U=0$.
Then  
$\Delta S$ and $\Delta T$ are defined as
\begin{align}
\Delta S&={16\pi} \frac{d}{dq^2}[\Pi_{33}(q^2)-\Pi_{3Q}(q^2)]|_{q^2\to0},\quad
\Delta T=\frac{16\pi}{s_{W}^2 m_Z^2}[\Pi_{\pm}(0)-\Pi_{33}(0)],
\end{align}
where $s_{W}^2\approx0.23$ is the Weinberg angle and $m_Z$ is the $Z$ 
boson mass, and $\Pi_{33(3Q)(\pm)}$ consists of the fermion-loop $L'$ and boson loop $\Phi_2$.
The fermion loop factors $\Pi_{33,3Q,\pm}^f(q^2)$ are calculated from the one-loop 
vacuum-polarization diagrams for $Z$ and $W^\pm$ bosons, which are respectively given by
\begin{align}
\Pi_{\pm}^f(q^2)&=
\frac{V^T_{\alpha+14,a} V^*_{a,\alpha+14}+V^\dag_{N_{\alpha+17,a}} V_{N_{a,\alpha+17}}}{(4\pi)^2}
\int_0^1 dx
\ln\left[-x(1-x)\frac{q^2}{M^2_{E'_\alpha}} + x + (1-x) \frac{M^2_{\psi_a}}{M^2_{E'_\alpha}}\right]
\nn\\&\times
\left[
2x(1-x)q^2 -x M^2_{E'_\alpha}- (1-x) M^2_{\psi_a}
\right],\\
\Pi_{33}^f(q^2)&=
\frac{1}{2(4\pi)^2}\int_0^1 dx
\left(
\ln\left[-x(1-x)\frac{q^2}{M^2_{E'_\alpha}}+ 1\right]
\left[2x(1-x)q^2 - M^2_{E'_\alpha} \right]\right. \\
&\left.+
\left[
(V^T_{N_{\alpha+14,a}} V^*_{N_{a,\beta+14}}) (V^T_{N_{\beta+14,b}} V^*_{N_{b,\alpha+14}})
+
(V^\dag_{N_{\alpha+17,a}} V_{N_{a,\beta+17}})(V^\dag_{N_{\beta+17,b}} V_{N_{b,\alpha+17}})
\right]
\right.\nn\\
&\left.\times
\ln\left[-x(1-x)\frac{q^2}{M^2_{\psi_a}} + x  + (1-x) \frac{M^2_{\psi_b}}{M^2_{\psi_a}}\right]
[2x(1-x)q^2 -x M^2_{\psi_a}- (1-x) M^2_{\psi_b}]\right),\nn\\
\Pi_{3Q}^f(q^2)&=
\frac{2}{(4\pi)^2}\int_0^1 dx
\ln\left[-x(1-x)\frac{q^2}{M^2_{E'_\alpha}}+ 1\right]
\left[2x(1-x)q^2 - M^2_{E'_\alpha} \right],
\end{align}
where $a(b)$ runs $1-20$, while $\alpha(\beta)$ runs $1-3$.
While the boson case are directly  given as $\Delta S^b$
 and $\Delta T^b$~\cite{Barbieri:2006dq};
 \begin{align}
& \Delta S^b\approx \frac{1}{2\pi}\int_0^1 dx x (1-x) \ln\left[\frac{x m^2_{h_2}+(1-x)m^2_{A^0}}{m^2_{H^\pm}}\right],\\
&\Delta T^b\approx \frac{1}{32\pi^2 v^2 \alpha_{em}}
[F(m_{H^\pm},m_{A^0}) + F(m_{H^\pm},m_{h_2}) -F(m_{A^0},m_{h_2}) ],\\
&F(m_{1},m_{2}) \equiv
\frac{m_1^2+m_2^2}{2}-\frac{m_1^2 m_2^2}{m_1^2 - m_2^2}\ln\frac{m^2_1}{m^2_2},
\end{align}
where we assume to be the no mixing among each of component $\Phi_2$. 
The experimental bounds are given by \cite{Patrignani:2016xqp}
\begin{align}
(0.07 - 0.08) \le \Delta S \le (0.07 + 0.08), \quad 
(0.10 - 0.07) \le \Delta T \le (0.10 + 0.07).
 \end{align}
In theoretical point of view, $\Delta T$ mainly corresponds to the mass differences between each of component inside the loop field, and $\Delta T=0$ is obtained in the limit of $M_{E'}=M_\Psi$ as well as $m_{h_2}=m_{H^\pm}$ without loss of generality. While $\Delta S$ corresponds to the number of new fields, and more new fields give more deviations from $\Delta S=0$. As another point of view, one can obtain opposite sign of contributions depending on the fermion loop or boson loop. For example, we always find positive value of $\Delta S^f$.
If the value of $\Delta S^f$ exceeds the experimental bound $0.15$, we can decrease the value by controlling $m_{h_2}< m_{H^\pm}$ whose condition leads to negative value of $\Delta S^b$.
As a quantitative aspect, the absolute value of $\Delta S$ is always less than 1 in our framework,
while the one of $\Delta T$ can fluctuate any value depending on the mass differences. Hence fitting the $\Delta T$
could be non-trivial and tends to be difficult.  
In addition, considering that $m_{h_2}$ can actually be considered as a free parameter and one can always be $\Delta S=0$, we focus on $\Delta T$.

\subsection{Collider Signatures}
\subsubsection{Issue of the Goldstone Boson}
 Here we show the mechanism that can generate a  nonzero mass for 
the Goldstone boson  $G$.
The mass is induced at higher order terms via gravitational effects 
that violate the global $U(1)_{}$ symmetry,
and its relevant Lagrangian is given by~\cite{Akhmedov:1992hi}
\footnote{These interactions among $G$ could affect invisible decays, 
cosmic string and so on.
However since these constraints are very weak due to the vector-like 
current~\cite{Nishiwaki:2015iqa, Hatanaka:2014tba}, 
we do not need to worry about these issues. See also, {\it i.e.}, Ref.~\cite{Cheung:2013oya} for discussing phenomenologies of GB at collider physics.}
\begin{align}
-\delta {\cal L}_G\sim \lambda_5 \frac{\varphi^5}{M_{pl}} +  \lambda_6 \frac{\varphi^4 \varphi^*}{M_{pl}} 
+ \lambda_7 \frac{\varphi^3 \varphi^{*2}}{M_{pl}} +{\rm c.c.},
\end{align}
where $M_{pl}\approx1.22\times10^{19}$ GeV is the Planck mass. From this 
dimension-5 operator, one straightforwardly finds the following mass for $G$:
\begin{align}
m_G\approx \frac{1}{5}\sqrt{\frac{25}{2}\lambda_5 + \frac{9}{2}\lambda_6+ \frac{1}{2}\lambda_7} \left[\frac{v'}{v_1}\right]^{3/2}\ {\rm keV},
\end{align}
where $v_1\ll v'$ is assumed. Here we suppose that the upper bound 
on $m_G$ is ${\cal O}$(1) MeV.

The Goldstone boson $G$ has the following interactions after the 
$U(1)_L$ symmetry breaking\cite{Weinberg:2013kea, Latosinski:2012ha}: 
\begin{align}
{\cal L}_{eff}
\supset
\frac{1}{v'}(\partial_\mu G) (\bar \ell \gamma^\mu \ell + \bar \nu \gamma^\mu P_L \nu).
\label{eq:int-GB}
\end{align}
Thus, we have annihilation modes of active neutrino pairs via $\varphi_R$  
in the s-channel.
This can be induced through the mixing among neutral fermions, 
and their interactions are found to be
\begin{align}
-{\cal L}_{\varphi\psi\bar\psi}&\sim
\frac{\varphi_R + i G}{\sqrt2} \left[(y_N)_{ij} \bar N_{R_{2i}} N_{L_{1j}} +  (y_N')_{kl} \bar N_{L_{1k}} N_{R_{0l}}+  (y_N'')_{mn} \bar N_{R_{1m}} N^C_{R_{0n}}\right]+{\rm h.c.}\nn\\
&=
(\varphi_R+ i G)
\sum_{a,b=1}^{20}\left[ (Y_L)^{ab}  \bar \psi_{a} {P_L} \psi_b^C + (Y_R)^{ab}  \bar \psi_{a}^C {P_R} \psi_b \right],
\end{align}
where
\begin{align}
(Y_L)^{ab} &= \frac1{\sqrt2}
\sum_{i, j=1}^3
\left[
 V^\ast_{N_{a,i+11}} (y_N)_{i,j} V^T_{N_{j+8,b}} 
-
V^\ast_{N_{a,i+8}} (y_N'^\dag)_{i, j} V^T_{N_{j+3,b}}
+
 V^\ast_{N_{a,i+5}} (y_N'')_{i, j} V^T_{N_{j+3,b}} \right] \nn\\
& =\sum_{i, j=1}^3
\frac{ V^\ast_{N_{a,i+11}} (M_{N_1})_{i,j} V^T_{N_{j+8,b} }
-
 V^\ast_{N_{a,i+8}} (M_{N_2}^\dag)_{i, j} V^T_{N_{j+3,b}}
+
V^\ast_{N_{a,i+5}} (M_{N_3})_{i, j} V^T_{N_{j+3,b}}}{v'}, 
\nn\\
(Y_R)^{ab} &= -(Y_L^\dag)^{ab},
 \label{GB-neut}
\end{align}
{where we have used the following relations: $VV^\dagger=1$ and $(\bar\psi_i\psi_j)^\dag=-(\bar\psi_j\psi_i)$.
Note that the GB interaction shown in Eq.~(\ref{eq:int-GB})
involves only the derivative couplings, which imply negligible 
contributions when coupled to vector currents.
}

\subsubsection{The scalar boson $\varphi_R$}

There are two relevant particles which may be of interests at colliders.
The first one is the $\varphi_R$, which mixes with $h_1$ through the mixing
angle $\alpha$. We have mentioned the current limit on $\alpha$ is
$\sin\alpha \alt 0.3$. Therefore, $\varphi_R$ could be produced in 
exactly the same ways as the SM Higgs boson, namely, dominated by the
gluon fusion (ggF) followed by vector-boson fusion, but suppressed by a factor
$\sin^2 \alpha \alt 0.09$.  For example, the ggF production rate for a
$\varphi_R$ of mass 500 GeV is approximately $5 \times 0.09 = 0.45$ pb. 
The decay modes for $\varphi_R$ are similar to those of the SM Higgs boson,
except that $\varphi_R \to H^0_1 H^0_1$ may now be possible and can be
dominant.  The size of this new channel depends on the 
$\lambda_{\varphi \Phi_1}$.

\subsubsection{Drell-Yan production of $E'^{+} E'^{-}$ }

The second relevant signature is through the Drell-Yan production of the
heavy charged fermion $E'^{-}_{L/R}$ of the doublet field $L'_{L/R}$. The 
$E'^{-}$ so produced will decay into the neutral component of the 
doublet and the $W$ boson (either real or virtual).  The $W$ boson can
decay into a charged lepton and a neutrino for leptonic detection. 
The neutral component $N'$ will decay into the SM neutrino(s) via mixing
and the SM Higgs boson(s).  One can detect the $b\bar b$ mode of the
Higgs decay. Therefore, the final state of $E'^{-} E'^{+}$ production
consists of two charged leptons and two  $b\bar b$ pairs at Higgs boson 
mass plus missing energies.

\section{Numerical analysis}
Here we randomly select points for the input parameters within the 
following ranges for both the cases of NH and IH:
\begin{align}
& v_2\in [0.1\,\text{}, 10]\ \text{GeV}, m_D\in [10^{-10}\,\text{}, 10^{-4}], \ m \in [10^{-4}\,\text{}, 50]\ \text{GeV},\nn\\
&
m_{H^\pm} \in [500\,\text{}, 1000]\ \text{GeV},
\quad m_{h_2}\in[0.01,10]\ \text{GeV},\quad m_{A^0}\in[0.01,100]\ \text{GeV},\nn\\
& [M_0, M_{N_1}, M_{N_2}, M_{N_3}, M_{D}, M, M', M_{L'}]\in [100\,\text{}, 1000]\ \text{GeV},
\end{align}
where {such a range of $m_{h_2}$ is taken in order to 
compensate for the fermion-loop contribution of $\Delta S^f$,
whose typical value is $0.5$}. 
In the last line, the range stands for all the elements for each matrix.
Note that $[M_{N_2}, M_{N_3}]$ are $3\times 2$ matrices,
$M_0$ is a $2 \times 2$ matrix, and $[M, M', M_{L'}]$ are 
$3\times 3$ diagonal matrices, 
while we assume that $[m, m_D, M_D, M_{N_1}]$ are $3\times 3$ 
symmetric matrices for simplicity.

\begin{figure}[t]\begin{center}
\includegraphics[width=0.80\columnwidth]{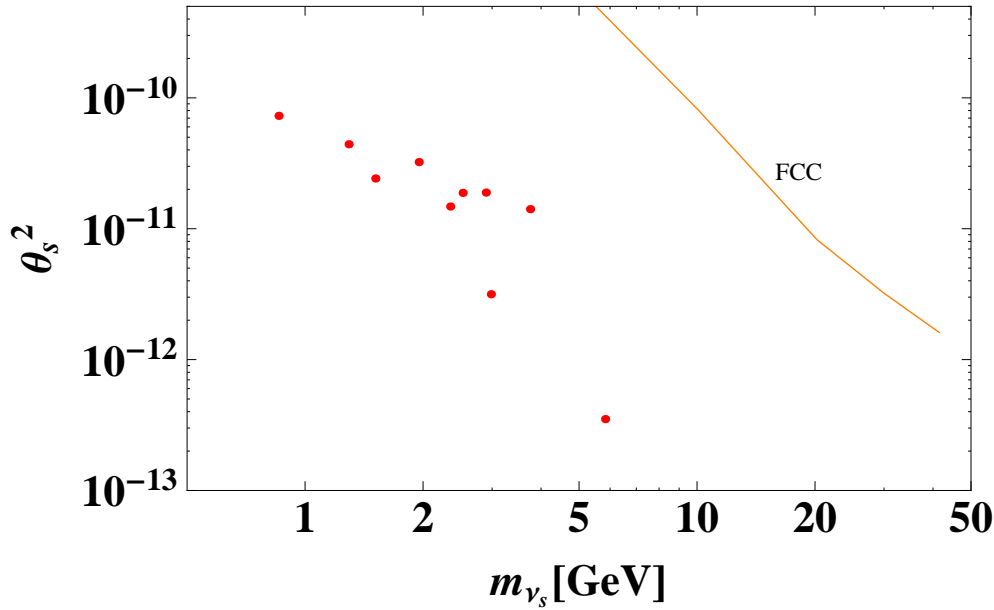}
   \caption{Plots in terms of $m_{\nu_s}$ and $\theta_s$ in case of NH,
where all the constraints discussed above are satisfied. The 
upper region {bounded by the orange line} 
is the favored region of detecting the sterile neutrino by FCC.}
   \label{dm-photon_NH}
\end{center}\end{figure}

\begin{figure}[t]\begin{center}
\includegraphics[width=0.80\columnwidth]{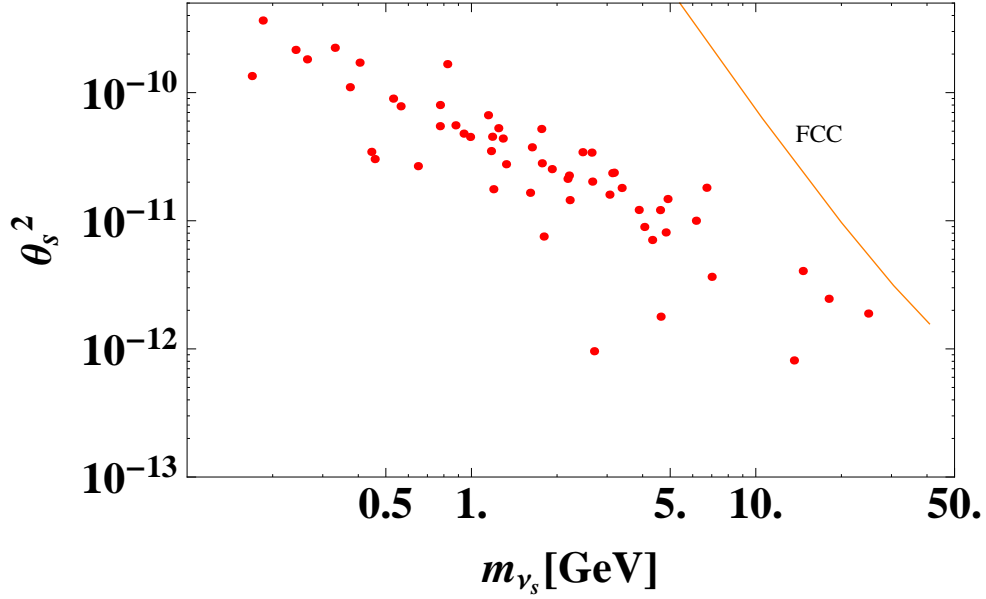}
   \caption{Plots in terms of $m_{\nu_s}$ and $\theta_s$ in case of IH,
where all the constraints discussed above are satisfied. The 
upper region {bounded by the orange line} 
is the favored region of detecting the sterile neutrino by FCC.}
   \label{dm-photon_IH}
\end{center}\end{figure}

We show scattered plots of {$\theta^2_s$ versus 
$m_{\nu_s}$ } in Fig.~\ref{dm-photon_NH} for NH and 
Fig.~\ref{dm-photon_IH} for IH.
The
red points satisfy the constraint of $\Delta T$. 
The allowed region ranges from $m_{\nu_s} \approx 0.5-50$ GeV with 
$\theta_s^2 \approx 10^{-12} - 5\times 10^{-10}$.
The lifetime of $\nu_s$ should be shorter than 0.1 second that is
equivalent to ${\cal O}(10^{23})$ GeV$^{-1}$. 
Notice here that, in addition to the usual modes such
as $\nu_s\to \ell W/\nu_L Z$ that appear in the canonical seesaw
scenario~\cite{Rasmussen:2016njh}, we also have the mode ${\nu_s}\to\nu_{L}
G$ via $Y_L/Y_R$, whose decay rate is given by
$\frac{m_{\nu_s}}{4\pi}\sum_{i=}^3|{\rm Im}(Y_L)^{4i}|^2$. 
Nevertheless, its lifetime is typically of the order $10^{-12}$ second
which is shorter than the standard decay modes. 
Thus, the BBN bound can be negligible.

The upper region bounded by the orange line is
covered in the FCC proposal, which gives the favored region 
of detecting the sterile neutrino in FCC.
We notice that in the region around 
$m_{\nu_s} \approx 20-50$ GeV and
$\theta_{s}^2 \approx 10^{-12}\sim 10^{-11}$,
the parameter space of our model is indeed covered by the 
FCC proposal for both cases of NH and IH.
{\it It implies that our testability with the FCC experiment is more
verifiable than the case of typical canonical seesaw
model, although the detector's luminosity should be improved to some extent.\footnote{In the typical canonical seesaw case, almost all of
the region can be tested by the FCC experiment~\cite{Rasmussen:2016njh}.}
This is the direct consequence of our huge matrix of the neutral fermions: $20\times 20$.}
Although the distribution of allowed region is similar between the NH
and IH cases, the number of IH solutions are larger than NH. This is a
natural consequence of the allowed range of neutrino oscillation data
in Eq.~(\ref{eq:exp-Neutmass-NH}) and Eq.~(\ref{eq:exp-Neutmass-IH}).

One might worry about the fact that we have no solution points that can
simultaneously satisfy the $\Delta T$ constraint and be covered by the FCC 
experiment.  
Also, $m_{h_2}$ may be too small to cause dangerous decays that violate our
scenario.  One of the simplest solutions is to introduce another boson
in isospin doublet.  For example, if we assign $(2,1/2,1/2)$ under
$(SU(2)_L, U(1)_Y, U(1)_L)$ for a new boson, we can obtain the measured
$\Delta T$ without violating our discussion above and its neutral
component can be a good dark matter candidate as an inert doublet
boson. Its mass is at around 500 GeV to satisfy the relic density,
which has already be discussed in Ref.~\cite{Hambye:2009pw}.

\section{ Conclusions and discussions}

We have proposed a model with two neutrinophilic Higgs doublet fields
$\Phi_{1,2}$, and the vacuum expectation value of the second Higgs
doublet is only induced at one-loop level. As a result, the active
neutrino masses can be naturally generated to be very small via the
tiny VEV $v_2$.  We have also discussed 
various phenomenology or constraints from neutrino
oscillation data, lepton-flavor violations, {the 
oblique parameters} and the muon $g-2$, 
and the possibilities of collider signatures.
In addition, we have pointed out a possibility of sterile neutrino of mass 
$O(0.1-10)$ GeV from the tiny VEV $v_2$.
Finally, we have shown a plot of $m_{\nu_s}$ and {$\theta^2_s$} 
that satisfy all the experimental bounds such as neutrino
oscillation data, LFVs, and the oblique parameters.
We have found an allowed region with 
$m_{\nu_s} \approx 20-50$ GeV and
$\theta_{s}^2 \approx 10^{-12}\sim 10^{-11}$ that is covered 
by the proposal of the future FCC in pursuing the sterile neutrinos.
{{\it It is one of the main results that our testability with the FCC 
experiment is more verifiable than the case of typical canonical 
seesaw model. This is the direct consequence of our huge matrix of 
the neutral fermions: $20 \times 20$.}}
For the muon $g-2$ we have obtained negative contributions that
seem to be against the experimental fact.  We may be able to detect
a signature by looking at the decay of $\varphi_R$ or by the Drell-Yan
process of $E'^+E'^-$ at the LHC.


At the end of the discussion, it is worthwhile to mention a new
possibility of detecting the Goldstone boson $G$.  According to a recent
work~\cite{Addazi:2017oge}, $G$ can be directly tested by the first
order phase transitions in the early Universe triggered by discovery
of gravitational waves at the experiment of
LIGO~\cite{Caprini:2015zlo}. All of the valid terms to explain it are
involved in our theory, our $G$ can also be tested near future.
 
\section*{Acknowledgments}
\vspace{0.5cm}
K.C. was supported by the MoST of Taiwan
under Grant No. MOST-105-2112-M-007-028-MY3.
H. O. is sincerely grateful for all the KIAS members in my stay.


\begin{appendix}
\section{Feynman Integrals}
Definitions of  $F\left(n, \alpha, \{ A_i; n\}\right)$ is the followings: 
\[
  F(n, \alpha, \{A_i; n\}) \equiv 
  \begin{cases}
\int_0^1 \prod_{i=1}^n dx_i \delta (1-\sum_{i=1}^n x_i)
\frac{1}{\left(\sum_{i=1}^n  x_i A_i\right)^\alpha}.  & (\alpha>0) 
\\
    \int_0^1 \prod_{i=1}^n dx_i \delta (1-\sum_{i=1}^n x_i)
\log \left(\sum_{i=1}^n  x_i A_i\right).  & (\alpha=0) 
\\
    \int_0^1 \prod_{i=1}^{n} dx_i \delta (1-\sum_{i=1}^n x_i)
 \left(\sum_{i=1}^{n}  x_i A_i\right)^{-\alpha}
\left(\log \left(\sum_{i=1}^{n}  x_i A_i\right)
- \sum_{i=1}^{-\alpha}\frac1i  \right).  & (\alpha<0)
  \end{cases}
\]
Where $\{ A_i; n\}=\{A_1, A_2, \cdots ,A_n\}$. 
The functions have following recurrence relations: 
\begin{eqnarray}
F(n, \alpha, \{A_i; n\})
&=&\frac{C_\alpha}{(A_{n-1}-A_n)} 
\nonumber\\ && \times
\left( F(n-1, \alpha-1, \{A_i; n-2, A_{n-1}\}) 
- F(n-1, \alpha-1, \{A_i; n-2, A_n\})\right), 
\nonumber\\
\label{eq:fnon0}
\end{eqnarray}
where $\{A_i; n-2, B\}=\{A_1, A_2, \cdots ,A_{n-2}, B\}$ and 
\[
  C_\alpha= 
  \begin{cases}
\frac{1}{1-\alpha}.  & (\alpha \neq1) 
\\
1.  & (\alpha =1) 
  \end{cases}
\]

We can obtain the formula of $F\left(n, \alpha, \{ A_i; n\}\right)$ using 
the recurrence relations and following relations:
\[
  F(1, \alpha, \{ A; 1 \}) = 
  \begin{cases}
\frac{1}{A^\alpha}.  & (\alpha>0) 
\\
\log \left(A\right).  & (\alpha=0) 
\\
 A^{-\alpha}
\left(\log A
- \sum_{i=1}^{-\alpha}\frac1i  \right).  & (\alpha<0)
  \end{cases}
  \label{eq:fn1}
\]

\end{appendix}

\end{document}